\documentclass{ws-ijmpd}

\newcommand{\ro}{r_0(\eta)}
\newcommand{\ep}{\epsilon}
\newcommand{\fin}{\phi_{in}}
\newcommand{\fo}{\phi_{out}}
\newcommand{\dte}{\frac{d\,t}{d\eta}}
\newcommand{\dre}{\frac{d\,r_0}{c\,d\eta}}
\newcommand{\mo}{M_{out}}
\newcommand{\mi}{M_{in}}

\begin{document}
\markboth{Belinski, Pizzi, Paolino}
{Charged membrane}

%
\catchline{}{}{}{}{}
%

\title{Charged membrane as a source for repulsive gravity}
\author{V. A. BELINSKI}

\address{INFN, Rome University ``La Sapienza'', 00185 Rome, Italy, \\
ICRANET, 65122, Pescara, Italy, and
                                                     \\
                                                                IHES, F-91440 Bures-sur-Yvette, France
\\
belinski@icra.it}

\author{M. PIZZI}

\address{Physics Department, Rome University ``La Sapienza'',\\
Piazzale A. Moro,  00185 Rome, Italy, and\\ ICRANet, 65122, Pescara, Italy.
\\
pizzi@icra.it}

\author{A. PAOLINO}

\address{Physics Department, Rome University ``La Sapienza'',\\
Piazzale A. Moro,  00185 Rome, Italy.}

\maketitle

\begin{history}
\received{Day Month Year}
\revised{08/08/2008}
\comby{Managing Editor}
\end{history}
\maketitle

\begin{abstract}
We demonstrate an alternative (with respect to the ones existing in literature) and more habitual for physicists derivation of exact solution of the Einstein-Maxwell equations for the motion of a charged spherical membrane with tangential tension. We stress that the physically acceptable range of parameters for which the static and stable state of the membrane producing the Reissner-Nordstrom (RN) repulsive gravity effect exists. The concrete realization of such state for the Nambu-Goto membrane is described. The point is that membrane are able to cut out the central naked singularity region and at the same time to join in appropriate way the RN repulsive region. 

As result we have a model of an everywhere-regular material source exhibiting a repulsive gravitational force in the vicinity of its surface: this construction gives a more sensible physical status to the RN solution in the naked singularity case.
\end{abstract}

\section{Introduction}
One of the interesting effects of relativistic gravity which has no analogue in the Newtonian theory is the presence of gravitational repulsive forces.
The classical example is the Reissner-Nordstrom (RN) field in the region close enough to the central singularity.
Indeed, in the RN metric
\begin{align}
-ds^{2}=-f\,c^2dt^2+f^{-1}\,dr^2 +r^2(d\theta^2+\sin^2\theta d\phi^2) \label{1} 
\end{align}
where
\begin{equation}\label{2}
      f=1-\frac{2kM}{c^2r}+\frac{kQ^2}{c^4r^2} \ ,
\end{equation}
the radial motion of a test neutral particle follows the equation:
\begin{equation}\label{3}
      \frac{d^2r}{ds^2}=-\frac{1}{2}\frac{df}{dr}=\frac{k}{c^4r^2}\left(\frac{Q^2}{r}-Mc^2 \right)
\end{equation}
from where one can see the appearance of repulsive force in the region of small $r$. In this zone the gradient of the gravitational potential $f(r)$ is negative and the gravitational force in Eq.(\ref{3}) is directed toward the outside of the central source.

For the RN naked singularity case ($Q^2>kM^2$), in which we are interested in the present paper, the potential $f(r)$ is everywhere positive and has a minimum at the point $r=Q^2/Mc^2$. Therefore at this point a neutral particle can stay at rest in the state of stable equilibrium (the detailed study can be found in [\refcite{1,Bon}] ). 

It is an interesting and nontrivial fact that the same sort of stationary equilibrium state due to the repulsive gravity exists also as an exact asymptotically flat two-body solution of the Einstein Maxwell equations which describes a Schwarzschild black hole situated at rest in the field of a RN naked singularity without any strut or string between these two objects [\refcite{AB,AB3}]. However, solutions of this kind have the feature that the object creating the repelling region has naked singularity and this last property has no clear physical interpretation. Consequently the pertinent question is whether the repelling phenomenon around a charged source arises only due to the presence of the naked singularity or it can be also a feature of physically reasonable structure of the space-time and matter.

By other words the question is whether or not it is possible to construct a regular material source which can block the central singularity and join the external repulsive region in a proper way. Then we are interested to construct a body with the following properties:

\begin{enumerate}
	\item inside the body there are no singularities;
	\item outside the body there is the RN field (\ref{1})-(\ref{2}), corresponding to the case $Q^2>kM^2$;
	\item the radius of the body is less than $Q^2/Mc^2$, so between the surface of the body and the sphere $r=Q^2/Mc^2$ arises the repulsive region;
	\item such stationary state of the body is stable with respect to collapse or expansion.
\end{enumerate}
In this paper we propose a new model for such body in the form of spherically symmetric thin membrane with positive tension. We assert that there exists a physically acceptable range of parameters for which all the above four conditions (1)-(4) can be satisfied. We illustrate this conclusion  by the especially transparent case of a Nambu-Goto membrane with equation of state $\ep=\tau$.

Then the existence of everywhere-regular material sources possessing RN ``antigravity'' properties in the vicinity of their surfaces attribute to this phenomenon and to the RN naked singularity solution more sensible physical status.

It is necessary to mention that at least two exact solutions of Einstein-Maxwell equations representing a compact continuous spherically symmetric distribution of charged matter under the tension producing the gravitationally repulsive forces inside the matter as well as in some region outside of it already exist in the literature. These are solutions constructed in Ref.[\refcite{5}] and Ref.[\refcite{6}]. A more detailed study of these two results can be found in Ref.[\refcite{7}]. An interesting possibility to have a gravitationally repulsive core of electrically neutral but viscous matter has been communicated in Ref.[\refcite{8}]. 

It is worth to remark that the first (to our knowledge) mentioning of the gravitational repulsive force due to the presence of electric field was made already in 1937 in the Ref.[\refcite{9}] in connection to the nonlinear model of electrodynamics of Born-Infield type. One of the first paper where a repulsive phenomenon in the framework of the conventional Einstein-Maxwell theory has been mentioned is Ref.[\refcite{DI}].
The general investigation of the different aspects of this phenomenon apart from the already mentioned references [\refcite{1}-\refcite{DI}] can be found also in the more detailed works [\refcite{11,Ips,12,13}]. Some part of these papers is dedicated to a possibility of construction a classical model for electron. This is doubtful enterprise, however, because the intrinsic structure of electron is a matter out of classical physics. Nonetheless the mathematical results obtained are useful and can be applied to the physically sensible situations, e.g. for construction the models of macroscopical objects.

\section{Equation of motion of a membrane with empty space inside}
\label{sec:EquationOfMotionOfAMembraneWithEmptySpaceInside}
The equation of motion for the most general case of a thin charged spherically symmetric fluid shell with tangential pressure moving in the RN field have been derived 38 years ago by J.E. Chase\cite{Cha}. The corresponding dynamics for a charged elastic membrane with tension follows from his equation simply by the change of the sign of the pressure.
We derived, however, the membrane's dynamics again using a different approach. 

Chase used the geometrical method which have been applied to the description of singular surfaces in relativistic gravity in [\refcite{Isr}] and have been elaborated in [\refcite{DI,16}] for some special cases of charged shells. An essential development of the Israel approach in application to the cosmological domain walls can be found in the series of works of V.Berezin, V.Kuzmin and I. Tkachev, see Ref.[\refcite{BKT}] and references therein.
Our treatment follows the method more habitual for physicists which have been used in [\refcite{BBB}], where the motion of a neutral fluid shell in a Schwarzschild field was derived by the direct integration of the Einstein equations with appropriate $\delta$-shaped source. Now we generalized this approach for the charged membrane and charged central source. 

Of course, the membrane's equation of motion that we obtained coincides with that of Chase. Nonetheless the different approach to the same problem often has a methodological value and gives new details. We hope that our case makes no exception, then for an interested reader we put the main steps of our derivation in Appendix (where we considered a 
 general case with central source).

In this section we study only the particular solution in which there is no central body, that is inside the membrane we have flat space-time.

Although the basic formulas of this section follow from the Appendix under restriction $M_{in}=Q_{in}=0$ the exposition we give here is more or less self-consistent. Only the definitions of 4-dimensional membrane's energy density and tension need some clarification which can be found in Appendix.

For the thin spherically symmetric membrane with empty space inside and with radius which depends on time the metrics inside, outside and on membrane are:

\begin{align}
	&-(ds^{2})_{in}=-\Gamma^2(t)c^2dt^2+dr^2 +r^2(d\theta^2+\sin^2\theta d\phi^2) \label{4} \\
  &-(ds^{2})_{out}=-f(r)c^2dt^2+f^{-1}(r) dr^2 +r^2(d\theta^2+\sin^2\theta d\phi^2) \label{5}\\
  &-(ds^{2})_{on}=-c^2d\eta^2+r_0^2(\eta)(d\theta^2+\sin^2\theta d\phi^2)  \label{6}
\end{align}

In the interval (\ref{6}) $\eta$ is the proper time of the membrane. The factor $\Gamma^2$ in (\ref{4}) is necessary to ensure the continuity of the global time coordinate $t$ through the membrane. The metric coefficient $f(r)$ in the region outside the membrane is given by Eq.(\ref{2}). 

Matching conditions for the intervals (\ref{4})-(\ref{6}) through the membrane's surface are:
\begin{align}
	[(ds^{2})_{in}]_{r=r_0(\eta)}=[(ds^{2})_{out}]_{r=r_0(\eta)}=(ds^{2})_{on} \label{7} 
\end{align}
If the equation of motion of the membrane $r=\ro$ is known, then from these conditions the connection $t(\eta)$ between global and proper times and factor $\Gamma(t)$ follow easily:
\begin{align}
	\Gamma(t)=\frac{f(r_0)\sqrt{1+c^{-2}(r_{0,\eta})^2}}{\sqrt{f(r_0)+c^{-2}(r_{0,\eta})^2}} \label{8} \\
	\dte=\frac{\sqrt{f(r_0)+c^{-2}(r_{0,\eta})^2}}{f(r_0)}    \label{9}	
\end{align}

The differential equation for the function $\ro$ follows from Einstein-Maxwell equations with energy-momentum tensor and charge current concentrated on the surface of the membrane. It is:
\begin{align}
	Mc^2=m(r_0) c^2\sqrt{1+\left(\dre\right)^2}+\frac{Q^2}{2r_0}-\frac{k\,m^2(r_0)}{2r_0} \label{10} 
\end{align}

Here $m(r_0)>0$ is the effective rest mass of the membrane in the radially comoving frame. This quantity includes the membrane's rest mass as well as all kinds of interaction mass-energies between membrane's constituents, that is those intrinsic energies which are responsible for the tension. The constants $Q$ and $M$ are the total charge of the membrane and total relativistic mass of the system. These are the same constants which appeared earlier in Eq.(\ref{2}).
The membrane's energy density $\ep$ and tension $\tau$ are (see Appendix for a further clarification):
\begin{align}
	\ep=\ep_0(r_0) \delta [r-r_0(\eta)]\, \ \ \ \tau=\tau_0(r_0) \delta[r-r_0(\eta)] \label{11} 
\end{align}
where
\begin{align}
&\ep_0 = \frac{m (r_0)c^2}{8\pi r_0^2}\left[\frac{1}{\sqrt{1+c^{-2}(r_{0,\eta})^2}}+\frac{f(r_0)}{\sqrt{f(r_0)+c^{-2}(r_{0,\eta})^2}}\right] \label{12} \\
&\tau_0(r_0)=\frac{dm(r_0)}{dr_0}\frac{r_0\ep_0(r_0)}{2m(r_0)} \label{13}
\end{align}

The electromagnetic potentials have the form $A_r=A_\theta=A_\phi=0$, $A_t=A_t(t,r)$ and for the electric field strength $\partial A_t/\partial r$ the solution is
\begin{equation}\label{14}
\frac{\partial A_t}{\partial r}=\left\{\begin{array}{lll}
\frac{Q}{r^2} & $for$ & r>r_0(\eta)\\\\
0 &      $for$ & r<\ro
\end{array}\right.  
\end{equation}
The formulas (\ref{4})-(\ref{14}) give the complete solution of the problem for the case of empty space inside the membrane.

Finally we would like to stress the following important point. As follows from discussion in Appendix, the signs of the square roots $\sqrt{1+c^{-2}(r_{0,\eta})^2}$ and $\sqrt{f(r_0)+c^{-2}(r_{0,\eta})^2}$ coincide with the signs of the time component $u^0$ of the 4-velocity of the membrane evaluated from inside and outside of the membrane respectively. The component $u^0$ is a continuous quantity by definition and can not change the sign when passing through the membrane's surface. Besides, for macroscopical objects we are interested in in this paper $u^0$ should be positive. Consequently the both aforementioned square roots should be positive. From another side it is easy to show that equation (\ref{10}) can be written also in the following equivalent form
\begin{align}
	Mc^2=m c^2\sqrt{f(r_0)+\left(\dre\right)^2}+\frac{Q^2}{2r_0}+\frac{k\,m^2}{2r_0} \label{15} 
\end{align}
Then from this expression and from (\ref{10}) follows that both square roots will be positive if and only if
\begin{align}
Mc^2-\frac{Q^2}{2r_0}-\frac{k\,m^2}{2r_0}>0 \label{16} 
\end{align}
This is unavoidable constraint which must be adopted as additional condition for any physically realizable solution of the equation of motion (\ref{10}) in classical macroscopical realm.

\section{Nambu-Goto membrane with ``antigravity'' effect}
\label{sec:NambuGotoMembraneWithAntigravityEffect}

To proceed further we must specify the function $m (r_0)$, which is equivalent to specifying an equation of state, as can be seen from (\ref{13}).

Let us analyze the membrane with equation of state $\ep=\tau$. This model can be interpreted as ``bare'' Nambu-Goto charged membrane\cite{KS,Hon}, or as Zeldovich-Kobzarev-Okun charged domain wall\cite{ZKO}. It follows from (\ref{13}) that for such type of membrane we have:
\begin{align}
m=\sigma r_0^2
\label{17} 
\end{align}
where $\sigma$ is an arbitrary constant. In this and next section we consider only the case of positive constants $\sigma$ and $M$:
\begin{align}
\sigma >0 \ , \ \ \ M>0\ .
\label{18} 
\end{align}
The sign of $Q$ is of no matter since the charge appear everywhere in square. Now we write the equation of motion (\ref{10}) in the following form:

\begin{align}
4\left(\dre\right)^2-\left(\frac{k\sigma r_0}{c^2}+\frac{2M}{\sigma r_0^2}-\frac{Q^2}{c^2\sigma r_0^3}\right)^2=-4\ .
\label{19} 
\end{align}
Formally this can be considered as the equation of motion of a non-relativistic particle with the ``mass'' equal to 8 moving in the potential $U(r_0)$,
\begin{align}
U(r_0)=-\left(\frac{k\sigma r_0}{c^2}+\frac{2M}{\sigma r_0^2}-\frac{Q^2}{c^2\sigma r_0^3}\right)^2
\label{20} 
\end{align}
and under that condition that particle is forced to live on the ``total energy'' level equal to minus four.

For the existence of the stable stationary state we are interested in, the following conditions should hold:
\begin{enumerate}
	\item The gravitational field in the exterior region should correspond to the super-extreme RN metric:
\begin{align}\label{21} 
Q^2>kM^2 .
\end{align}

\item The potential $U(r_0)$ should have a local minimum at some value $r_0=R_{min}$. The form (\ref{20}) of $U(r_0)$ permit this if and only if
\begin{align}\label{22}
 k\sigma^2 Q^6<(Mc^2)^4 .
\end{align}
Under this restriction the potential $U(r_0)$ has three extrema, two maxima at points $r_0=R_{max}^{(1)}$ and $r_0=R_{max}^{(2)}$ and a minimum which is located between them: $R_{max}^{(1)}<R_{min}<R_{max}^{(2)}$. We show the shape of the potential $U(r_0)$ for this case in Fig.1. 
\begin{figure}
\includegraphics[width=11cm, height=11cm]{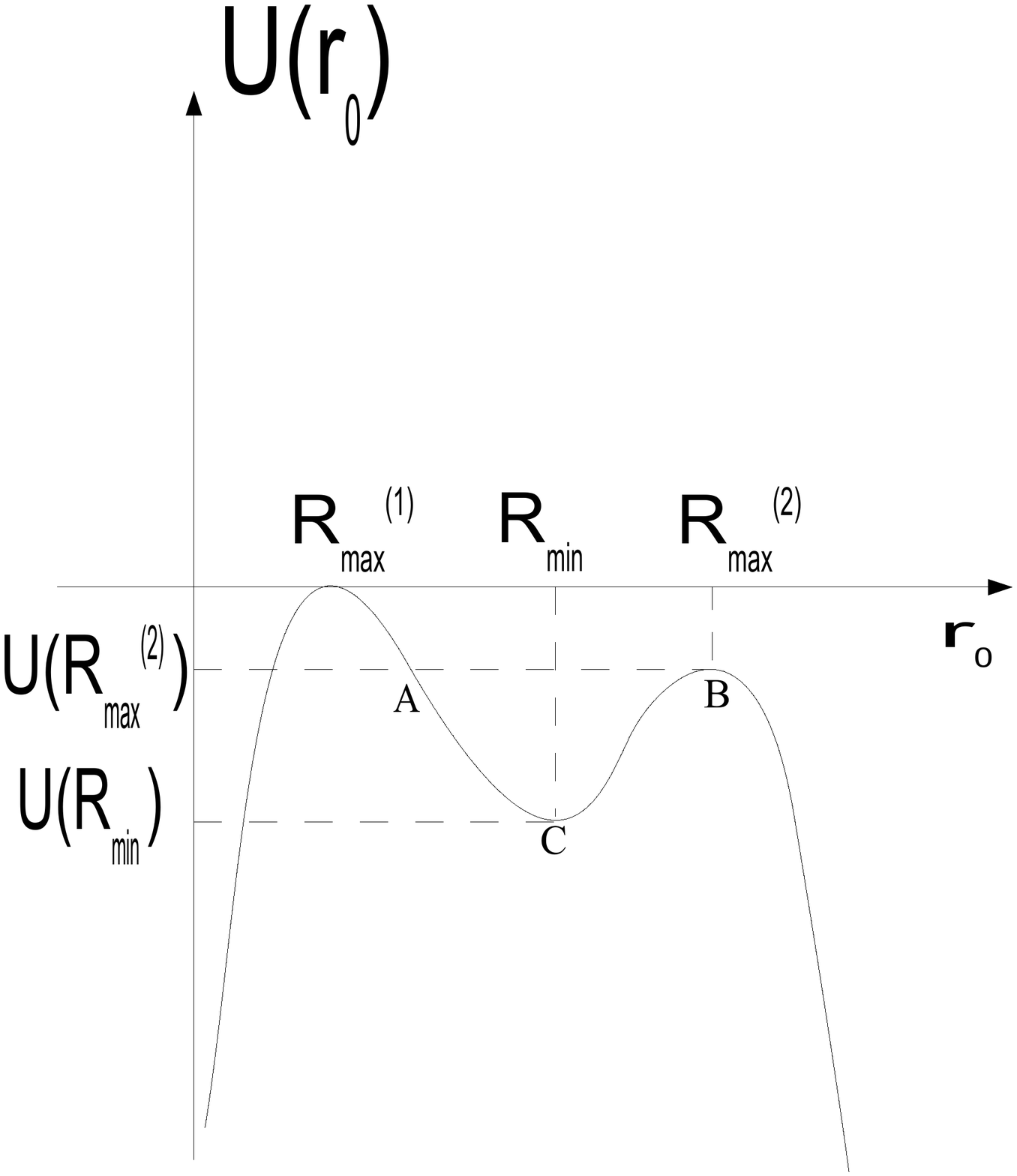}
  \caption{The membrane's motion can be described as the motion of a non-relativistic point particle in the potential $U(r_0)$.}\label{f.1}
\end{figure}

\hspace{1cm}The equation $U(r_0)=0$ has only one real root and this is also the first local maximum $R_{max}^{(1)}$. The minimum and the second maximum are coming as two other roots of the equation $\frac{dU}{dr_0}=0$.

The equation for $R_{min}$ is:
\begin{equation}\label{23}
 k\sigma^2 R_{min}^4-4Mc^2 R_{min}+3 Q^2=0 \ .
\end{equation}
This fourth order equation has only two real solutions and $R_{min}$ is the smaller one.

\item  For the stationary position of the membrane at the minimum of the potential we must ensure the relation $U(R_{min})=-4$ which is:

\begin{align}\label{24}
\frac{k\sigma}{c^2}R_{min} +\frac{2M}{\sigma}R_{min}^{-2}-\frac{Q^2}{c^2\sigma}R_{min}^{-3}=2
\end{align}
(the minus two in the r.h.s. of (\ref{24}) would be incompatible with Eq.(\ref{23}) under condition (\ref{18})).

\item To have repulsive region it is necessary for the membrane's radius $R_{min}$ to be less than the minimum of the gravitational potential $f(r)$, that is less than the quantity $Q^2/Mc^2$. In this case outside of the membrane surface in the region $R_{min}<r<Q^2/Mc^2$ we have the repulsive effect. Then we demand:

\begin{align}\label{25}
R_{min}<\frac{Q^2}{Mc^2} \ .
\end{align}

 \item Also the additional constraint (\ref{16}) should be satisfied. 
 This means that for our stationary solution we have to satisfy the inequality:
\begin{align}\label{26}
Mc^2-\frac{Q^2}{2R_{min}}-\frac{k\sigma^2}{2}R_{min}^3>0 \ .
\end{align}

\item We have also another condition: that the electric field nearby the membrane should be not too large, otherwise the stability of the model would be destroyed by the strong macroscopical consequences of quantum effects, e.g. by the intensive electron-positron pair creation.
This condition (which was suggested by J.A. Wheeler long time ago, see the reference with this Wheeler's proposal in the paper of Bekenstein\cite{18}) is:
\begin{align}\label{27}
&\frac{Q}{R_{min}^2}<<\mathcal E_{cr}\ , & \mathcal E_{cr}=\frac{m_e^2c^3}{e_e\hbar}\ ,
\end{align}
where $m_e$ and $e_e$ are the electron's mass and charge). $\mathcal E_{cr}$ is the well known critical electric field above which the intensive process of pair creation starts.
\end{enumerate}
To satisfy these six conditions we have to find a physically acceptable domain in the space of the four parameters $M$, $Q$, $\sigma$ and $R_{min}$. The point is that such domain indeed exists and it is wide enough. If we introduce the dimensionless radius of the stationary membrane $x$ as 
\begin{align}\label{28}
\frac{k\sigma}{c^2}R_{min}=x \ ,
\end{align}
then one can check directly that the first five of the above formulated conditions will be satisfied under the following three constraints:
\begin{align}
&x<1  \label{29} \\
&M=\frac{c^4}{k^2\sigma}(3x^2-2x^3) \label{30}\\
&Q^2=\frac{c^8}{k^3\sigma^2}(4x^3-3x^4) \label{31}
\end{align}
The last two of these relations are just the equations (\ref{23}) and (\ref{24}) but written in the form resolved with respect to $M$ and $Q^2$.

The formulas (\ref{29})-(\ref{31}) shows that for the first five conditions it is convenient to take $x<1$ and $\sigma$ as independent parameters, and then to calculate the mass and charge necessary to obtain the model we need. 

As for the last constraint (\ref{27}) it gives some restriction also for parameter $\sigma$:
\begin{align}\label{32}
k\sigma^2<<\frac{x}{4-3x}\mathcal E_{cr}^2\ .
\end{align}
The energy density $\ep$ for the stationary state at $r_0=R_{min}$, expressed in terms of parameters $x$ and $\sigma$, is:
\begin{align}\label{33}
\ep=\frac{\sigma c^2}{8\pi}(1+\sqrt{x^2-2x+1})\delta(r-R_{min}).
\end{align}

\section{Summary}
\label{sec:su}

\hspace{0.5cm}1. We showed that exists a possibility to have a spherically charged membrane in stable stationary state producing RN repulsive gravitational force outside its surface and having flat space inside. To construct such model one should take a pair of constants $0<x<1$ and $\sigma >0$ satisfying the inequality (\ref{32}) and calculate from (\ref{28}) and (\ref{30})-(\ref{31}) the membrane's radius $R_{min}$, total mass $M$ and charge $Q$.

2. The equation of motion (\ref{10}) can be used also for the description of the oscillation of the membrane in the potential well ABC (see fig.1) above the equilibrium point C. If we slightly increase the total membrane's energy $Mc^2$ then the potential $U(r_0)$ around its minimum (i.e. the point C and its vicinity) will be shifted slightly down but the level "minus four'' in Eq.(\ref{20}) on which the system lives will remain at the same position. Then the membrane will oscillate between the new shifted walls AC and CB.

3. It is easy to see that in the general dynamical state the membrane can live only inside the potential well ABC. All regions outside ABC are forbidden. In the region to the right from the point $R_{max}^{(2)}$ and above the potential $U(r_0)$ any location of the membrane is impossible due to the fact that inequality (\ref{16}) is violated there.

This means that a membrane of considered type in principle can not have the radius (no matter in which state) greater than $R_{max}^{(2)}$. In turn for $R_{max}^{(2)}$ it is easy to obtain from the potential (20) the upper limit $R_{max}^{(2)}<\frac{c^2}{k\sigma} \left(\frac{4 k^2\sigma M}{c^4}\right)^{1/3}$. 

The same violation of the inequality (\ref{16}) take place in the domain between $R_{max}^{(1)}$ and $R_{max}^{(2)}$ and above the segment AB. The motion in the region to the left from the point $R_{max}^{(1)}$ and above the curve $U(r_0)$ is forbidden again due to the same violation of the condition (\ref{16}).
This means that a membrane of considered type in principle can not have the radius less than $R_{max}^{(1)}$. 
In particular there is no way for a membrane with positive effective rest mass $m$ to collapse to the point $r_0=0$ leaving outside the field corresponding to the RN naked singularity solution. This conclusion is in agreement with the main result of the paper [\refcite{19}].

4. Although we claimed that the stationary state of a membrane constructed is stable this stability should be understood in a very restrict sense, that is as stability in the framework of the dynamics described by the equation (\ref{10}). We do not know what will happen to our membrane after the whole set of arbitrary perturbations will be given.

5. In general the arbitrary perturbations will change also the equation of state. We investigated a membrane with equation of state $\ep=\tau$. However this case can be considered only as ``bare'' Nambu-Goto membrane, by other words as a toy model. In the papers [\refcite{KS,Hon,Pol,Car90,Vil,HKS}] it was shown that arbitrary perturbations essentially renormalize the form of the equation of state of the strings and membranes. Moreover for the membranes [\refcite{Hon}] (differently from the strings) the fixed points of the renormalization group for the transverse and longitudinal perturbations does not coincide, which means that for the general ``wiggly'' membrane there is no equation of state of the type $\ep=\ep(\tau)$ at all.

6. We also would like to stress that for appearance of repulsive force the presence of electric field is of no principal necessity. For example the repulsive gravitational forces arise also in neutral viscous fluid \cite{8} and in the course of interaction between electrically neutral topological gravitational solitons \cite{Bel91}.

7. From the conditions (21)-(26) also follows that in addition to the inequality (25) the radius $R_{min}$ of the shell in the stable stationary state cannot be less than $\frac{Q^2}{2Mc^2}$. A simple analysis shows that there is no way for $R_{min}$ to be arbitrarily small keeping some finite non-zero value for $M$ and $Q$.
\section*{Appendix}
\label{sec:Appendix}
For the spherically symmetric case the metric\footnote{We use the 
notations in which the interval is written as $-ds^2=g_{ik} dx^idx^k $ and metric signature is $(-,+,+,+)$, i.e. the time-time component $g_{00}$ is negative. The norm of a time-like vector is negative. The Roman indices take values $0,1,2,3$. The Newtonian constant is denoted by $k$. The simple partial derivatives we designated by a comma, while covariant derivatives by semicolon.} is:
\begin{align}
-(ds)_{in}^{2}=g_{00} c^2dt^2+g_{11} dr^2 +r^2(d\theta^2+\sin^2\theta d\phi^2) \label{a1}
\end{align}
where $g_{00}$ and $g_{11}$ depend only on $t,\ r$ and the standard notation for the coordinates is:
\begin{align}\label{a2}
(x^0,x^1,x^2,x^3)=(ct,r,\theta,\phi)\ .
\end{align}
The Electromagnetic tensor $F_{ik}$ has the form:
\begin{align}
F_{ik}=A_{k,i}-A_{i,k}\label{a3}
\end{align}
and the Einstein-Maxwell equations are:
\begin{align}
 R_i^k-\frac{1}{2}R \delta_i^k=\frac{8\pi k}{c^4}T_i^k   \label{a4}\\
        (F^{ik})_{;k}=\frac{4\pi}{c} \rho u^{i}   \label{a5}
\end{align}
The energy-momentum tensor for a spherical charged membrane with energy density $\ep$ and tangential tension $\tau$ is:
\begin{align}\label{a6}
T_i^k=\epsilon \,u_i u^k-(\delta_i^2\delta_2 ^k+\delta_i^3\delta_3^k)\tau+\frac{1}{4\pi}(F_{il}F^{kl}-\frac{1}{4} \delta_i^kF_{lm}F^{lm})
\end{align}
and for the membrane's 4-velocity $u^i$ we have:
\begin{align}\label{a7}
u^0=u^0(t,r), \ u^1=u^1(t,r), \ u^2=u^3=0 \ , \ \ \ \ \ u^iu_i=-1 \ .
\end{align}
The main step is to define the 4-invariant charge and energy densities $\rho$ and $\ep$. After that, the tension $\tau$ follows  automatically from the Einstein-Maxwell equations and from the equation of state. To construct $\rho$ and $\ep$  we apply the Landau-Lifschitz procedure\cite{LL}. 

The charge $dq$ in the 3-volume element $dV=\sqrt{g_{11}g_{22}g_{33}}\,dx^1dx^2dx^3$ is a 4-invariant quantity by definition 
 (although $dV$ is not a 4-scalar).
 The three-dimensional charge density $\rho^{(3)}$ can be introduced by the relation $dq=\rho^{(3)}dV$. Consequently, for the spherically symmetric membrane case it is:
\begin{align}\label{a7b}
\rho^{(3)}=\frac{Q\delta(r-r_0)}{4\pi r^2\sqrt{g_{11}}} \ ,
\end{align}
where $Q$ is the electric charge of the membrane and $r_0$ is the membrane's radius. Indeed it is easy to check that $Q=\int\rho^{(3)}dV$ as it should be\footnote{The $\delta$-function in curved metric (\ref{a1}) is defined by the usual relation $\int \delta(r-r_0)dr=1$. Such $\delta$-function has dimension $cm^{-1}$.}.

Since $\rho^{(3)}dV$ is a 4-scalar the quantities $\rho^{(3)}dVdx^i$ represent a 4-vector. With the use of the previous formula we obtain:
\begin{align}\label{a7c}
c\rho^{(3)}dVdx^i=\frac{c \,Q\delta(r-r_0)}{4\pi r^2u^0\sqrt{-g_{00}g_{11}}}u^i\sqrt{-g} \,d^4x \ ,
\end{align}
where $g$ is the 4-metric's determinant. The last formula shows that the factor in front of $u^i\sqrt{-g}d^4x$ is a 4-scalar. This scalar is nothing else but the 4-invariant charge density $\rho$ which appeared in the Maxwell equation (\ref{a5}):
\begin{align}\label{a8}
\rho=\frac{cQ\delta[r-r_0(t)]}{4\pi r^2u^0\sqrt{-g_{00}g_{11}}}.
\end{align}
For the electric current $j^k$ we have $j^k=\rho u^k$.

The 4-scalar energy density $\ep$ which figure in the energy-momentum 4-tensor (\ref{a6}) can be constructed exactly in the same way if we observe that the rest energy of the matter in a 3-volume element $dV$ (i.e. the sum of the all kinds of the internal energies of this element in the reference system in which this element is at rest) is a 4-invariant quantity by definition. Then we can introduce the 3-dimensional rest energy density (the direct analogue of the previous charge density $\rho^{(3)}$) which under integration over 3-volume gives the total rest energy $m c^2$ of the membrane. Then $m c^2$ is the sum of the all kinds of internal energies of the membrane in the radially comoving system in which membrane is at rest. In this way we obtain:
\begin{equation}	\label{a9}
\ep=\frac{m c^2\delta[r-r_0(t)]}{4\pi r^2 u^0\sqrt{-g_{00}g_{11}}}
 \ .
\end{equation}
Clearly  the effective rest mass $m$ of the membrane in the presence of a tension depends on the membrane radius $r_0(t)$.

In the case of spherical symmetry the electromagnetic potentials $A_i$ can be taken in the form:
\begin{equation}	\label{a10}
A_0=A_0(t,r),\ \ A_1=A_2=A_3=0,
\end{equation}
which gives only one nonvanishing component for the electromagnetic tensor $F_{ik}$, namely $F_{10}$ (and its antisymmetric partner $F_{01}$):
\begin{align}\label{a11}
F_{10}=A_{0,1}  \ .
\end{align}
Now, we enter with definitions (\ref{a1})-(\ref{a3}) and (\ref{a6})-(\ref{a11}) into the Einstein-Maxwell equations (\ref{a4})-(\ref{a5}) to calculate the solution. 
These calculations need special care since we are dealing with distributions in application to the non-linear theory. In general this is not a trivial task (see e.g. [\refcite{Tau,GT,SV}]), however, for particular case of spherical symmetry everything is tractable and can be done easily thanks to the specially simple structure of the field equations. The resulting solution contains four arbitrary constants of integration $M_{in}$, $Q_{in}$ and $M_{out}$, $Q_{out}$ which have an obvious interpretation as mass and charge of a central RN source and the total mass and charge of the whole system (the central body together with the membrane) respectively. The membrane's charge $Q$ is simply the difference of $Q_{out}$ and $Q_{in}$:
\begin{align}\label{a12}
Q=Q_{out}-Q_{in} \ .
\end{align}
To represent the solution in compact form we use the proper time $\eta$ of the membrane, denoting the membrane's equation of motion as $r=r_0(\eta)$, and introducing the following notations:
\begin{align}\label{a13}
 \left.\begin{array}{c}
	\fin(r)=1-\frac{2k\,M_{in}}{c^2r}+\frac{kQ_{in}^2}{c^4r^2}\\\\
	\fo(r)=1-\frac{2k\,M_{out}}{c^2r}+\frac{kQ_{out}^2}{c^4r^2}\end{array}\right\}
\end{align}
\begin{align}
\left. \begin{array}{c}
	S_{in}(\eta)=\sqrt{\fin(r_0)+c^{-2}(r_{0,\eta})^2}\\\\
	S_{out}(\eta)=\sqrt{\fo(r_0)+c^{-2}(r_{0,\eta})^2}\end{array}\right\}\label{a14}
\end{align}

We consider the global time $t$ in (\ref{a1}) as continuous quantity when passing through the membrane. Then the intervals inside, outside and on the membrane are:
\begin{align}
	&-(ds^{2})_{in}=-\Gamma^2(t)\fin(r) c^2dt^2+\frac{dr^2}{\fin(r)} +r^2(d\theta^2+\sin^2\theta d\phi^2) \label{a15} \\
  &-(ds^{2})_{out}=-\fo(r)c^2dt^2+\frac{dr^2}{\fo(r)} +r^2(d\theta^2+\sin^2\theta d\phi^2) \label{a16}\\
  &-(ds^{2})_{on}=-c^2d\eta^2+r_0^2(\eta)(d\theta^2+\sin^2\theta d\phi^2)  \label{a17}
\end{align}
The matching conditions for these intervals  through the membrane are:
\begin{align}\label{a18}
[(ds^{2})_{in}]_{r=r_0(\eta)}=[(ds^{2})_{out}]_{r=r_0(\eta)}=(ds^{2})_{on}
\end{align}
Using the relations (\ref{a18}), the factor $\Gamma(t)$ in (\ref{a15}) and the connection $t(\eta)$ between global and proper times can be expressed through the membrane's radius $r_0(\eta)$:
\begin{align}\label{a19}
\frac{dt}{d\eta}=\frac{S_{out}}{\fo(r_0)} \ ,  \\
\Gamma (t)=\frac{\fo(r_0)S_{in}}{\fin(r_0)S_{out}} \ . \label{a20}
\end{align}
Namely the continuity conditions (\ref{a18}) and continuous character of the time variable $t$ are responsible for the appearance of the term $\Gamma^2(t)$ in $g_{00}$ in Eq.(\ref{a19}). Since this term depends only on time, it can be easily removed by passing to the internal time variable $t_{in}$ by the transformation
\begin{align}\label{a21}
\Gamma dt=dt_{in} \ ,
\end{align}
which can be found with the help of (\ref{a19}) after the function $r_0(\eta)$ became known. In terms of the variables $(t_{in},\,r)$ also the internal metric (\ref{a15}) takes the standard RN form.

As it was already mentioned, the membrane's effective rest mass $m$ which appeared in the energy density (\ref{a9}) depends on the membrane radius. The concrete form of the function $m(r_0)$ is not known in advance and its specification is equivalent to the specification of the equation of state. For an arbitrary $m(r_0)$ the Einstein-Maxwell equations (\ref{a4})-(\ref{a5}) give the following equation of motion for the membrane:
\begin{align}\label{a22}
\mo c^2-\mi c^2=\frac{1}{2}(S_{in}+S_{out})m c^2+\frac{QQ_{in}}{r_0}+\frac{Q^2}{2r_0} \ ,
\end{align}
together with the condition that both square roots $S_{in}$ and $S_{out}$ defined by (\ref{a14}), should have the same sign.
The provenance of this condition is due to the fact that the signs of $S_{in}$ and $S_{out}$ are nothing else but the signs of the time-component of $u^0$ of the membrane's 4-velocity when it is seen from the inside ($r\rightarrow r_0-0$) and outside ($r\rightarrow r_0+0$) of the membrane surface respectively. In our approach (with continuous coordinates $t,r$) we can consider the 4-velocity $u^i$ as a field continuous through the surface of the membrane. We can define $u^i$ everywhere in space-time simply by smooth parallel transport from the membrane's surface, no matter that the membrane is concentrated only at the points $r=r_0$. This concentration is ensured not by $u^i$ but due to the $\delta$-functions in the densities $\rho$ and $\ep$. Since $u^0$ can not change sign passing through the membrane, $S_{in}$ and $S_{out}$ should have the same sign.

Of course, we need to know the fields $u^0$ and $u^1$ only on the membrane, and there they are:
\begin{align}\label{a23}
u^0=t_{,\eta} \ \ \ \  ; \ \ \ \  u^1=c^{-1}r_{0,\eta}
\end{align}
It is easy to check that the matching conditions (\ref{a19}) and (\ref{a20}) are nothing else but the demand that the normalization constraint $u^iu_i=-1$ should hold independently from which side we approach the surface of the membrane.

It is worth to be remarked that the Einstein-Maxwell equations also demand for the trajectory $r_0(\eta)$ the second order (in time) differential equation of motion. However, this last one represents simply the result of the differentiation in time of the first order equation (\ref{a22}). Then this second-order equation we can forget safely.

The resulting expressions for the energy density and tension are:
\begin{align}
&\ep=\frac{m c^2}{8\pi r_0^2}\left[\frac{\fin(r_0)}{S_{in}}+\frac{\fo(r_0)}{S_{out}}\right]\delta[r-r_0(\eta)]   \label{a24}\\
&\tau=\frac{r_0}{2m}\frac{dm}{dr_0}\ep \ . \label{a25}
\end{align}
The electric field $F_{10}$ outside the membrane is:
\begin{align}\label{a26}
F_{10}=\frac{Q_{out}}{r^2}\ , \ \ \ r>r_0 \ .
\end{align}
Inside the membrane we have:
\begin{align}\label{a27}
F_{10}=\frac{Q_{in}}{r^2}\frac{dt_{in}}{dt}\ , \ \ \ r<r_0 \ ,
\end{align}
where the factor $\frac{dt_{in}}{dt}$ depends only on time and can be calculated from the relations (\ref{a20}) and (\ref{a21}). The origin of this factor is due to the fact that we use the time $t$ as continuous global time including the region inside the membrane. If we describe the internal metric in terms of internal time $t_{in}$ the field strength $F_{10}$ would be simply $Q_{in}/r^2$. 

The formulas (\ref{a13})-(\ref{a17}), (\ref{a19}), (\ref{a20}) and (\ref{a22})-(\ref{a27}) provide the complete solution of the problem. It is worth explaining briefly the main steps of our integration procedure that we applied to the Einstein-Maxwell equations.

As in any spherically symmetric problem it is convenient to use, instead of the full original Einstein equations (\ref{a4}), only its $(_0^0)$, $(_1^1)$ and $(^1_0)$ components, and the hydrodynamical equations $T^k_{i;k}=0$. All the remaining components of equations (\ref{a4}) after that will be satisfied identically either due to the Bianchi identities or due to the symmetry of the problem. Then the solution for $g_{11}$ together with the basic eq.(\ref{a22}) follows from $(_0^0)$ and $(_0^1)$ components of Einstein equations (\ref{a4}), and after that the solution for $g_{00}$ follows from the difference of the $(_0^0)$ and $(_1^1)$ components of (\ref{a4}).
The solution for the electric field $F_{10}$ is the result of the Maxwell equations (\ref{a5}). 
The hydrodynamical equations $T^k_{i;k}=0$ give only two relations. The first one simply express the tension $\tau$ in terms of other quantities and this is the formula (\ref{a25}). The second one results in the already mentioned second order differential equation for $r_0(\eta)$ which represents the differentiation in time of the first order equation (\ref{a22}). Then this second order equation is of no importance.

We remark also  that the procedure described above need a caution because the symbolic function are involved. Nevertheless everything going well under the following three standard operation rules with such functions:
\begin{enumerate}
	\item $\frac{d}{dx}\theta(x)=\delta(x)$ ,\\ 
	\item $F(x)\delta(x)=\frac{1}{2}[F(-0)+F(+0)]\delta (x)$ , \\ 
	\item $\frac{d}{dx}\theta^2(x)=2\theta(x)\delta(x)=\delta(x)$ .
\end{enumerate}
(To call the third rule as the standard one is a little exaggeration; however it works well and final results indeed coincide with those obtained in literature by different approaches).
Originally we obtained the solution in global form using the step function $\theta(x)$ and only after that we represented the results separately in the regions $r>r_0$ and $r<r_0$. However, since $\theta(x)$ is defined also at the point $x=0$ [$\theta(0)=1/2$], we found by the way the values for the metric and electric field also at the points of the membrane's surface. Such global form is:
\begin{align}
& \frac{1}{g_{11}}=1-\frac{2k\mi }{c^2 r}-\frac{2k(\mo-\mi)}{c^2r}\theta[r-r_0(\eta)]+\frac{k}{c^4 r^2}\left\{Q_{in}+Q\theta[r-r_0(\eta)] \right\}^2  \\
& \frac{1}{\sqrt{-g_{00}g_{11}}}=\frac{1}{\Gamma}+\left(1-\frac{1}{\Gamma}\right)\theta[r-r_0(\eta)] \\
& F_{10}=\frac{\sqrt{-g_{00}g_{11}}}{r^2}\left\{Q_{in}+Q\theta[r-r_0(\eta)] \right\} \ ,
\end{align}
to which should be added the equation (\ref{a22}). This equation arise as self-consistency condition for the $(_0^0)$ and $(_0^1)$ components of Einstein equations, which can be verified by the direct substitution into these components of the above global expressions together with eqs. (\ref{a23})-(\ref{a25}). 

Finally it should be mentioned that the membrane's equation of motion (\ref{a22}) can be written in the following two equivalent forms:
\begin{align}
& m c^2 S_{in}=\mo c^2-\mi c^2-\frac{Q_{in}Q}{r_0}-\frac{Q^2}{2r_0}+\frac{k\,m^2}{2r_0}  \label{a28} \\
& m c^2 S_{out}=\mo c^2-\mi c^2-\frac{Q_{in}Q}{r_0}-\frac{Q^2}{2r_0}-\frac{k\,m^2}{2r_0} \ .  \label{a29}
\end{align}
Each of these two equations is equivalent to (\ref{a22}) which can be checked easily by simple algebraic manipulations. For practical calculations we can use only one of these equations, however, in addition it is necessary to ensure the same sign for both quantities $S_{in}$ and $S_{out}$. (For a membrane with empty space inside they both should be positive). More convenient is relation (\ref{a28}) which we write as 
\begin{align}
\mo c^2=&\mi c^2+m c^2\sqrt{\fin(r_0)+c^{-2}(r_{0,\eta})^2} \nonumber \\
        & +\frac{Q_{in}Q}{r_0}+\frac{Q^2}{2r_0}-\frac{k\,m^2}{2r_0} \label{a30}
\end{align}
This is the equation obtained by Chase\cite{Cha} with the aid of a different derivation procedure which makes use of Gauss-Codazzi equations (see Israel\cite{Isr}).

Eqn.(\ref{a30}) is interesting because in spite of the fact that $m$ depends on time (or on $r_0$) this equation looks like an usual integral of motion, that is as if $m$ was a constant. Relation (\ref{a30}) expresses the conservation of the total energy  $M_{out}c^2$ of the system which is the sum of the five familiar constituents: 1) the rest energy of the central body, 2) the kinetic energy of the membrane together with its gravitational potential energy in the gravitational field of the central body, 3)the electric interaction energy between membrane and central source, 4) the positive electric self-interaction energy of the membrane, and 5) the negative gravitational self-interaction energy of the membrane.

\bibliographystyle{unsrt}
\bibliography{thin_shell}

\end{document}